\documentclass[journal]{IEEEtran}
\usepackage{mathrsfs}
\usepackage{pifont}
\usepackage{bbding}
\usepackage{tipa}
\usepackage{amsmath,epsfig}
\usepackage{stmaryrd}
\usepackage{amssymb}
\usepackage{amsfonts}
\usepackage{epic}
\usepackage{graphicx}
\usepackage{curves}
\usepackage{mathbbold}
\ifCLASSINFOpdf
\else
\fi
\begin{document}
\newtheorem{theorem}{Theorem}
\newtheorem{proposition}{Proposition}
\newtheorem{definition}{Definition}
\newtheorem{lemma}{Lemma}
\newtheorem{corollary}{Corollary}
\newtheorem{remark}{Remark}
\newtheorem{construction}{Construction}

\newcommand{\supp}{\mathop{\rm supp}}
\newcommand{\sinc}{\mathop{\rm sinc}}
\newcommand{\spann}{\mathop{\rm span}}
\newcommand{\essinf}{\mathop{\rm ess\,inf}}
\newcommand{\esssup}{\mathop{\rm ess\,sup}}
\newcommand{\Lip}{\rm Lip}
\newcommand{\sign}{\mathop{\rm sign}}
\newcommand{\osc}{\mathop{\rm osc}}
\newcommand{\R}{{\mathbb{R}}}
\newcommand{\Z}{{\mathbb{Z}}}
\newcommand{\C}{{\mathbb{C}}}
%

\title{Design of network-coding based multi-edge type LDPC codes for multi-source relaying systems}
\author{\IEEEauthorblockN{Jun~Li, Marwan~H.~Azmi, Robert.~Malaney, and Jinhong Yuan}\IEEEauthorblockA{\\School of Electrical Engineering and Telecommunications,\\
University of New South Wales, Sydney, NSW 2052, AUSTALIA\\ Email:
jun.li@unsw.edu.au,~marwan@student.unsw.edu.au,~r.malaney@unsw.edu.au,~jinhong@ee.unsw.edu.au}}


%

\maketitle

\begin{abstract}
In this paper we investigate a multi-source LDPC scheme for a Gaussian
relay system, where $M$ sources communicate with the destination under the
help of a single relay ($M-1-1$ system). Since various distributed LDPC
schemes in the cooperative single-source system, e.g. bilayer LDPC~\cite{IEEEconf:1} and
 bilayer multi-edge type LDPC (BMET-LDPC)~\cite{IEEEconf:2}, have been designed to approach the Shannon limit,
these schemes can be applied to the $M-1-1$ system by the relay serving each source in a
round-robin fashion. However, such a direct application is not optimal
due to the lack of potential joint processing gain. In this paper, we propose a network coded
multi-edge type LDPC (NCMET-LDPC) scheme for the multi-source scenario.  Through an EXIT analysis,
we conclude that the NCMET-LDPC scheme achieves higher extrinsic mutual information, relative to
 a  separate application of BMET-LDPC to each source. Our new NCMET-LDPC scheme
thus achieves a higher threshold relative to existing schemes.
\end{abstract}



%
\IEEEpeerreviewmaketitle

\section{Introduction}\label{sec:1}
Low-density parity-check (LDPC) codes have been shown to approach theoretical capacity limits for
single link communication channels~\cite{IEEEconf:1}. Recently,
distributed LDPC for cooperative communications has attracted much attention. The work of \cite{IEEEconf:2} first explored the
the use of bilayer LDPC codes within the cooperative single source channel ($1-1-1$ system),
where full-duplexing relay is used. Although bilayer LDPC is carefully designed to
approach the system capacity~\cite{IEEEconf:4}, the performance is decreased
as the capacity gap between the source-to-relay channel and the
source-to-destination channel becomes larger. In~\cite{IEEEconf:5}, multi-edge type
LDPC code has been utilized to address  this problem~\cite{IEEEconf:3}. The works of
~\cite{IEEEconf:6,IEEEconf:7,IEEEconf:8} consider more practical issues in the $1-1-1$ system,
such as the use of Rayleigh fading channels and half-duplexing relays.

\begin{figure}[!t]
\centering
\includegraphics[width=3in,angle=0]{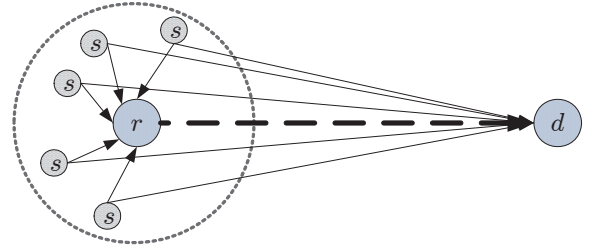}
\caption{Multi-source relay system.}\label{fig1}
\end{figure}

However, the above studies on distributed LDPC codes are all
limited to the triangle model, which contains only one source. In
this paper, we investigate a network coding~\cite{IEEEconf:9}
based LDPC codes designed for the cooperative uplink system with multi-source and one relay ($M-1-1$
system) as shown in Fig.~\ref{fig1}. Based on existing methods of
distributed LDPC design in the triangle model, an intuitive thought
is that the relay serves the sources in a round-robin fashion,  optimizing the
distributed LDPC for a single source in each round. Unfortunately, such a direct application is not
optimal for the following reasons.
\begin{enumerate}
\item{The check digits produced by the relay in the $m$-th, $(m=1,\cdots,M)$, round are only
based on the codeword from the source $s_m$, which is highly
correlated with the original check codes already produced by $s_m$.}
\item{Network coded check digits enable the joint decoding of all sources' data and thus each source can
obtain more extrinsic mutual information from the other sources.}
\item{The code profile optimization executed by the
relay aims only at approaching the capacity for the single source,
which leads to a sub-optimal outcome for a multi-source
system in terms of network capacity.}
\end{enumerate}

We note that~\cite{IEEEconf:10} has studied  the network coded LDPC in a multi-source system
with fading channels. But the code design in ~\cite{IEEEconf:10} is not optimal.
~\cite{IEEEconf:11} also proposes a joint bilayer LDPC scheme in the $M-1-1$
system. However, this scheme is constrained by the drawbacks of
bilayer LDPC. So it cannot deal with the problem where the capacity
gap between source-to-relay channel and source-to-destination is
large. Also,~\cite{IEEEconf:11} only considers a scenario
where all channel capacities in the model are equal. In this paper, we
propose a network coding based multi-edge type LDPC, which we refer to as NCMET-LDPC, for the $M-1-1$ model which addresses these issues.
In our analysis of NCMET-LDPC we utilize extrinsic mutual information (EMI) transfer,
rather than density evolution (DE) used in~\cite{IEEEconf:11}, which aids in understanding the code design issues better.
Through the EXIT chart, we see that the network coded parity check digits produced by relay provide more EMI for each
source, relative to a separate round-robbin scheme.
\section{System Model and Preliminaries}\label{sec:2}
Consider a Gaussian relay system with $M$-sources, $1$-relay and $1$
destination as shown in Fig.~\ref{fig1}, where sources $s_1, \cdots, s_M$
transmit information to the destination $d$ simultaneously with the help of
a full-duplex relay $r$. We assume all sources are randomly distributed around the
relay. Suppose that the $m$-th source, $s_m$,
transmits its information in the frequency band $f_m$, and $r$ can
receive and transmit at $f_1, \cdots, f_M$. All the frequency
bands are assumed to be orthogonal. With these constraints the multi-source system can
be viewed as $M$ independent parallel $1-1-1$ systems. Within each of these channels, a
bilayer LDPC scheme~\cite{IEEEconf:2} can be utilized to approach the $1-1-1$ system
capacity~\cite{IEEEconf:4}.

\begin{figure}[!t]
\centering
\includegraphics[width=3.5in,angle=0]{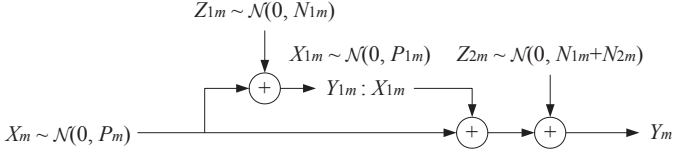}
\caption{The Gaussian relay system working at frequency band $f_m$:
$X_m$ is power constrained to $P_m$; $X_{1m}$ is power constrained to
$P_{1m}$.}\label{fig2}
\end{figure}

Let us briefly review the achievable rate and code design in the $1-1-1$
system~\cite{IEEEconf:2}. Without loss of generality, we focus firstly on a specific $1-1-1$ system. This is formed by selecting a specific value of $m$,
and constructing a model composed of $s_m, r_m$ and $d$, where $r_m$ is the part of
$r$ operating at $f_m$. In Fig.~\ref{fig2}, $X_m$ is the signal transmitted by
$s_m$, which has the average power $P_m$, and $X_{1m}$ is the signal transmitted
by $r_m$, which has the average power $P_{1m}$. The binning scheme is used to
achieve the capacity in a Gaussian degraded relay channel~\cite{IEEEconf:4}. In this binning scheme,
$s_m$ divides its total power $P_m$ into a fraction $\alpha{P_m}$ for the new codeword
$\omega_i$, and a fraction $(1-\alpha)P_m$ for the bin index $\phi_i$ of the previous
codeword $\omega_{i-1}$. So in the $i$-th time slot, $X_m$ is the superposition of
$\omega_i$ and $\phi_i$, which will be received by both $r_m$ and $d$.
Since $r_m$ has successfully decoded $\omega_{i-1}$ in the previous time slot, $\omega_{i}$
will be successfully decoded at the relay with a rate no more than
\begin{equation}\label{equ:1}
R_+^m=\frac1{2}\log\left(1+\frac{\alpha{P_m}}{N_{1m}}\right).
\end{equation}
Meanwhile, $r_m$ is transmitting $X_{1,m}(\phi_i)$ to $d$. Thus,
$d$ receives the interferential signal composed by $X_m$ and $X_{1m}$ in
the frequency band $f_m$. By successive interference cancellation,
$d$ firstly treats $X_m$ as noise so as to extract the bin index $\phi_i$
with a rate no more than
\begin{equation}\label{equ:2}
R_1^m=\frac1{2}\log\left(1+\frac{(\sqrt{P_{1m}}+\sqrt{(1-\alpha)P_{m}})^2}{\alpha{P_m}+N_{1m}+N_{2m}}\right).
\end{equation}
Then combining with $\phi_i$, the decoding of $\omega_{i-1}$ at $d$ will be successful with a rate no more than
\begin{equation}\label{equ:3}
R_-^m=\frac1{2}\log\left(1+\frac{\alpha{P_m}}{N_{1m}+N_{2m}}\right).
\end{equation}
Combining the above three equations, we get the overall rate for the Gaussian relay channel working
at $f_m$ as
\begin{equation}\label{equ:4}
R_m=\underset{\alpha}{\max}\min\{R_+^m,R_1^m+R_-^m\}.
\end{equation}
So to maximize $R_m$, we let $R_+^m=R_1^m+R_-^m$ by adapting power allocation $\alpha$.

For the practical code design, we can utilize the bilayer LDPC scheme
for each $s_m~(m=1,\cdots,M)$ to approach $R_+^m$ and $R_-^m$~\cite{IEEEconf:2}.
For example, the code structure is divided into the lower
and the upper graphs and consists of $k_{1m}$ and $k_{2m}$ check nodes, respectively.
Both type of check nodes are connected to the same $n_m$ variable nodes. According to~\cite{IEEEconf:2}, in the
design of such codes, one should firstly determine the optimal LDPC code corresponding to the lower graph
to achieve rate $R_+^m$, and then search over the whole bilayer graph in order to find a good LDPC code that approaches the rate $R_-^m$.
So in the multi-source case, a straightforward technique of practical code design
is to perform bilayer LDPC individually to each source.
As previously mentioned, such individual processing does not benefit from any joint processing gain at the relay.
In the following section, we focus on the network coded multi-edge type LDPC code for the multi-source system.
\section{Multi-edge Type LDPC for Multi-source System}\label{sec:3}
\subsection{Multi-edge Type LDPC Codes}
The principle of multi-edge type LDPC is to introduce more than one edge type to the Tanner
graph~\cite{IEEEconf:5}, where the graph ensemble is specified through two polynomials, one
associated to variable nodes and the other associated to constraint nodes. The two polynomials
are given by
\begin{equation}\label{equ:5}
v(\textbf{r},\textbf{x})=\boldsymbol{\sum}v_{\textbf{b},\textbf{d}}\textbf{r}^{\textbf{b}}\textbf{x}^{\textbf{d}}\quad
\text{and}\quad{\mu}(\textbf{x})=\boldsymbol{\sum}\mu_{\textbf{d}}\textbf{x}^{\textbf{d}},
\end{equation}
where $\textbf{d}=(d_1,\cdots,d_{n_l})$ is a multi-edge degree and $\textbf{x}=(x_1,\cdots,x_{n_l})$ denotes variables.
Similarly, $\textbf{b}=(b_0,\cdots,b_{n_\tau})$ is a received degree, and $\textbf{r}=(r_0,\cdots,r_{n_\tau})$
denotes variables corresponding to received distributions. We use $\textbf{x}^{\textbf{d}}$ to denote $\prod_{i=1}^{n_l}x_i^{d_i}$
and $\textbf{r}^{\textbf{b}}$ to denote $\prod_{i=0}^{n_\tau}r_i^{b_i}$. For more details about multi-edge type
LDPC codes one can refer to~\cite{IEEEconf:5}.
\begin{figure}[!t]
\centering
\includegraphics[width=3in,angle=0]{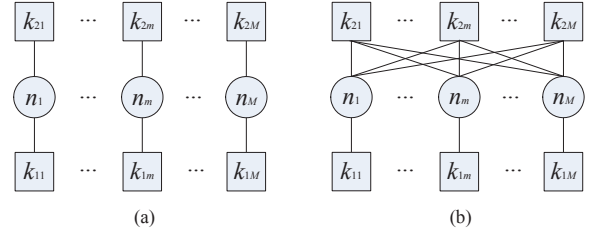}
\caption{Comparison of the two LDPC code schemes. (a) The conventional bilayer LDPC code design for each source individually;
(b) The proposed multi-edge type LDPC with joint processing.}\label{fig4}
\end{figure}
\subsection{Design of Multi-edge Type LDPC}
\begin{figure}[!t]
\centering
\includegraphics[width=1.5in,angle=0]{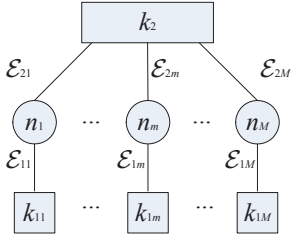}
\caption{Tanner graph and multi-edge type in proposed scheme.}\label{fig5}
\end{figure}

In this subsection, we propose a novel multi-edge type LDPC scheme in the $M-1-1$ system. In the new coding
scheme, we do not change the power allocation of each transmitter (i.e network capacity is not altered), but propose a new coding scheme to approach the network capacity.
Fig.~\ref{fig4} compares the proposed scheme with the conventional bilayer LDPC code.
In the conventional scheme, the additional $k_{2m}$ check digits at $r_m$ are only produced from $s_m$'s frame and
transmitted at $f_m$. However, in the proposed scheme, the extra $k_{2m}$ bits of information are co-produced from
all the sources' frames and randomly distributed in all the frequency bands. $r$ jointly processes all sources' information
and produces $k_2=\boldsymbol\sum_{m=1}^Mk_{2m}$ parity check digits, which can be seen as a super parity check block.
The Tanner graph of the multi-edge type LDPC is shown as Fig.~\ref{fig5}, where we represent the edge types
as $\mathcal {E}$ with different subscripts. The edge of the lower graph
of $s_m$ is denoted as $\mathcal {E}_{1m}$ and the edge of the upper graph is
denoted as $\mathcal {E}_{2m}$. We also assume the frames from all the
sources have the same length, i.e., $n_1=\cdots=n_M=n$.

The multi-edge type LDPC code design for an $M-1-1$ system begins with optimizing the lower Tanner graph in Fig.~\ref{fig5}
at rate $R_+^1,\cdots,R_+^M$ for $s_1,\cdots,s_M$, respectively, which follows the conventional methodology of
single link LDPC codes. So the lower graph ensemble of the multi-edge type LDPC codes for $s_m$ is represented by
\begin{equation}\label{equ:6}
\begin{split}
v_m(\textbf{r},\textbf{x})&=r_{1}\underset{d_{1,m}=1}{\overset{d_{v1,m}}{\boldsymbol{\sum}}}v_{[0,1],[d_{1,m}]}x_{1,m}^{d_{1,m}},
\\\mu_m(\textbf{x})&=\underset{d_{1,m}=1}{\overset{d_{c1,m}}{\boldsymbol{\sum}}}\mu_{[d_{1,m}]}x_{1,m}^{d_{1,m}},
\end{split}
\end{equation}
where $[0,1]$ is the vector $\textbf{b}$, and $[d_{1,m}]$ is the vector $\textbf{d}$ of (\ref{equ:5}).
For vector $\textbf{b}$, all variable nodes in the codeword are transmitted through the source-to-relay channel
$(b_1=1)$ at rate $R_+^m$ and there are no punctured variables $(b_0=0)$ in the codeword.
Vector $\textbf{d}$ contains only one element since there is only one edge type. $d_{1,m}$
is denoted as the degree of variable nodes and check nodes with the maximum value $d_{v1,m}$ and $d_{c1,m}$, respectively.
The quantity $v_{\textbf{b},\textbf{d}}n$ is the the number of variable nodes of type $(\textbf{b},\textbf{d})$ and $\mu_{\textbf{d}}n$
is the number of check nodes of the type $\textbf{d}$ in the graph.
The code rate of the lower graph ensemble for $s_m$ is
\begin{equation}\label{equ:7}
R_+^{m}=1-\underset{d_{1,m}=1}{\overset{d_{c1,m}}{\boldsymbol\sum}}\mu_{[d_{1,m}]}.
\end{equation}
So the sum of code rate for the lower graph ensembles for all the sources is
\begin{equation}\label{equ:8}
R_+=\underset{m=1}{\overset{M}{\boldsymbol\sum}}R_+^m=M-\underset{m=1}{\overset{M}{\boldsymbol\sum}}
\underset{d_{1,m}=1}{\overset{d_{c1,m}}{\boldsymbol\sum}}\mu_{[d_{1,m}]}.
\end{equation}

In the next step, we design the overall graph considering the relay jointly processing all of the sources' information.
Relay $r$ will transmit additional $k_2=\boldsymbol\sum_{m=1}^Mk_{2m}$ parity check digits to $d$ within the
relay-to-destination channel capacity $\boldsymbol\sum_{m=1}^MR_1^m$,
in which $R_1^mn$ bits are allocated to $s_m$. So the variable and check nodes' polynomials for all sources
$s_m$ in the overall graph can be written as
\begin{equation}\label{equ:9}
v_m(\textbf{r},\textbf{x})=r_1\underset{d_{1,m}=1}{\overset{d_{v1,m}}{\boldsymbol\sum}}
\underset{d_{2,m}=0}{\overset{d_{v2,m}}{\boldsymbol\sum}}v_{[0,1],[d_{1,m},d_{2,m}]}x_{1,m}^{d_{1,m}}x_{2,m}^{d_{2,m}}\quad\:
\end{equation}
\begin{equation*}
\mu_m(\textbf{x})=\underset{d_{1,m}=1}{\overset{d_{c1,m}}{\boldsymbol\sum}}\mu_{[d_{1,m},0,\cdots,0]}x_{1,m}^{d_{1,m}}+
\underset{d_{2,m}=1}{\overset{d_{c2,m}}{\boldsymbol\sum}}\mu_{[0,d_{2,1},\cdots,d_{2,M}]}x_{2,m}^{d_{2,m}} .
\end{equation*}
These relations mean that for the overall graph ensemble, all variable nodes in the codeword are transmitted through the source-to-destination
channel $(b_1=1)$ at rate $R_-^m$, and there are no punctured variables $(b_0=0)$. Vector $[d_{1,m},d_{2,m}]$ in
(\ref{equ:9}) represents the number of the sockets of the two edge types $\mathcal {E}_{1m}$ and $\mathcal {E}_{2m}$ in a
variable node with $d_{1,m}$ and $d_{2,m}$ respectively. Vector $[d_{1,m},0,\cdots,0]$ in
(\ref{equ:9}) represents the number of the sockets of the edge type $\mathcal {E}_{1m}$ in a check node of the lower graph
with $d_{1,m}$. Since the check nodes in the lower graph are only connected to $\mathcal {E}_{1m}$, the number of the sockets
of other edge types is all zero. Vector $[0,d_{2,1},\cdots,d_{2,M}]$ in the second equation of
(\ref{equ:9}) represents the number of the sockets of the edge types $\mathcal {E}_{21},\cdots,\mathcal {E}_{2M}$ in a check node of the upper graph
with $d_{2,1},\cdots,d_{2,M}$. Since the check nodes in the upper graph are not connected to $\mathcal {E}_{1m}$, the number of the sockets
of this edge type is zero. Note that for $s_m$ we have
\begin{equation}\label{equ:10}
R_-^m\le1-\underset{d_{1,m}=1}{\overset{d_{c1,m}}{\boldsymbol\sum}}\mu_{[d_{1,m},0,\cdots,0]}-
\underset{d_{2,m}=1}{\overset{d_{c2,m}}{\boldsymbol\sum}}\:\:\underset{\sim d_{2,m}}{\boldsymbol\sum}\mu_{[0,d_{2,1},\cdots,d_{2,M}]},
\end{equation}
where
\begin{equation}\label{equ:11}
\begin{split}
\underset{\sim d_{2,m}}{\boldsymbol\sum}&=\underset{d_{2,1}=0}{\overset{d_{c2,1}}{\boldsymbol\sum}}\cdots\
\underset{d_{2,m-1}=0}{\overset{d_{c_2,m-1}}{\boldsymbol\sum}}\:\:
\underset{d_{2,m+1}=0}{\overset{d_{c_2,m+1}}{\boldsymbol\sum}}\cdots
\underset{d_{2,M}=0}{\overset{d_{c_2,M}}{\boldsymbol\sum}}.
\end{split}
\end{equation}
Since the $k_2$ parity check digits are shared by $M$ types of edges, their contribution to $s_m$ is
\begin{equation}\label{equ:12}
R_{1}^m=\underset{d_{2,m}=1}{\overset{d_{c2,m}}{\boldsymbol\sum}}\:\:
\underset{\sim d_{2,m}}{\boldsymbol\sum}\frac{\mu_{[0,d_{2,1},\cdots,d_{2,M}]}d_{2,m}}{\boldsymbol\sum_{l=1}^{M}d_{2,l}}.
\end{equation}
Then we get
\begin{equation}\label{equ:13}
R_-^m=1-\underset{d_{1,m}=1}{\overset{d_{c1,m}}{\boldsymbol\sum}}\mu_{[d_{1,m},0,\cdots,0]}-R_1^m.
\end{equation}
So the code rate of the overall graph ensemble can be computed as follows.
\begin{equation}\label{equ:14}
R_-=M-\underset{m=1}{\overset{M}{\boldsymbol\sum}}\left(\underset{d_{1,m}=1}{\overset{d_{c1,m}}{\boldsymbol\sum}}\mu_{[d_{1,m},0,\cdots,0]}+R_1^m\right).
\end{equation}

We optimize the whole system by regarding all the sources' frames as a super block,
which accesses the relay with code rate $R_+$ for the lower graph, and accesses  the destination with
code rate $R_-$ for the overall graph. Besides (\ref{equ:7}), (\ref{equ:12}) and (\ref{equ:13}),
there are several constraints that should be satisfied for each $s_m$ as follows;
\begin{equation}\label{equ:15}
\begin{split}
&v_{[0,1],[d_{1,m}]}=\underset{d_{2,m=0}}{\overset{d_{v2,m}}{\boldsymbol\sum}}v_{[0,1],[d_{1,m},d_{2,m}]},\\&\mu_{[d_{1,m}]}=\mu_{[d_{1,m},0,\cdots,0]}
\\&\underset{d_{1,m=1}}{\overset{d_{c1,m}}{\boldsymbol\sum}}\mu_{[d_{1,m}]}d_{1,m}=\underset{d_{1,m=1}}{\overset{d_{v1,m}}{\boldsymbol\sum}}v_{[0,1],[d_{1,m}]}d_{1,m}
\\&\underset{d_{2,m}=1}{\overset{d_{c2,m}}{\boldsymbol\sum}}\:\:\underset{\sim d_{2,m}}{\boldsymbol\sum}\mu_{[0,d_{2,1},\cdots,d_{2,M}]}d_{2,m}=\\&\qquad\qquad\qquad\qquad
\underset{d_{1,m=1}}{\overset{d_{v1,m}}{\boldsymbol\sum}}\:\:\underset{d_{2,m=0}}{\overset{d_{v2,m}}{\boldsymbol\sum}}v_{[0,1],[d_{1,m},d_{2,m}]}d_{2,m}.
\end{split}
\end{equation}
To deduce the average extrinsic mutual information of each edge type, we characterize the code ensemble of $s_m$ by the
degree distribution $\lambda_{[d_{1,m},d_{2,m}]}^{{i,m}}$, for $i=1,2$,
\begin{equation}\label{equ:16}
\lambda_{[d_{1,m},d_{2,m}]}^{{i,m}}=\frac{v_{[0,1],[d_{1,m},d_{2,m}]}d_{i,m}}{\boldsymbol\sum_{d_{1,l=1}}^{d_{v1,m}}
\boldsymbol\sum_{d_{2,l=0}}^{d_{v2,m}}v_{[0,1],[d_{1,l},d_{2,l}]}d_{1,l}}.
\end{equation}
 This defines the percentage of $\mathcal {E}_{im}$ type edges connected to the
variable nodes with $d_{1,m}$ edges in $\mathcal {E}_{1m}$,
and $d_{2,m}$ edges in $\mathcal {E}_{2m}$.
We also define another two types of degree distribution. One is $\rho_{[d_{1,m}]}^{1,m}$, which denotes the percentage of $\mathcal {E}_{1m}$ type edges
connected to the check nodes in the lower graph with $d_{1,m}$ edges, i.e.,
\begin{equation}\label{equ:17}
\rho_{[d_{1,m}]}^{1,m}=\frac{\mu_{[d_{1,m}]}d_{1,m}}{\boldsymbol\sum_{d_{1,l=1}}^{d_{c1,m}}\mu_{[d_{1,l}]}d_{1,l}},
\end{equation}
and the other is $\rho_{[0,d_{2,1},\cdots,d_{2,M}]}^{2,m}$, which denotes the percentage of $\mathcal {E}_{2m}$ type edges connected to the check nodes with
the edge vector $[0,d_{2,1},\cdots,d_{2,M}]$, i.e.,
\begin{equation}\label{equ:18}
\rho_{[0,d_{2,1},\cdots,d_{2,M}]}^{2,m}=\frac{\mu_{[0,d_{2,1},\cdots,d_{2,M}]}d_{2,m}}{\boldsymbol\sum_{d_{2,l}=0}^{d_{c2,m}}
\boldsymbol\sum_{\sim d_{2,l}}\mu_{[0,d_{1},\cdots,d_{M}]}d_{2,l}}.
\end{equation}
\section{Performance Analysis and Code Design}\label{sec:4}
In~\cite{IEEEconf:2}, Density Evolution (DE) is applied to the bilayer LDPC code profile optimization.
Due to the fixed degree of the check nodes in both the lower graph and the upper graphs, the complexity of DE is tolerable.
However, the fixed degree deteriorates the system performance as mentioned in the introduction.
So in the optimization of the multi-edge type LDPC code, we exploit the extrinsic information transfer (EXIT) functions~\cite{IEEEconf:12,IEEEconf:13,IEEEconf:14}
to reduce the code searching complexity. This will likely lead to better code profiles by canceling the constraints
on the check nodes degree as in bilayer LDPC code.

We denote the variable nodes set associated with the codeword bits of $s_m$ as $V_m$, and the check nodes set associated with
the parity check digits of $s_m$ from the lower graph as $C_{1,m}$. The shared check nodes set in the upper graph is denoted as $C_{2}$.
Since $V_m$ is connect to two edge types, i.e., $C_{1,m}$ and $C_{2}$, there are four types of mutual information (MI) defined as follows~\cite{IEEEconf:14}.

$I_{Ev}(1,m)$: The MI between the message sent from $V_m$ to $C_{1,m}$ and the associated codeword bit, on each edge in the edge type $\mathcal {E}_{1m}$ connecting $V_m$ to $C_{1,m}$.

$I_{Ev}(2,m)$: The MI between the message sent from $V_m$ to $C_{2}$ and the associated codeword bit, on each edge in the edge type $\mathcal {E}_{2m}$ connecting $V_m$ to $C_{2}$.

$I_{Ec}(1,m)$: The MI between the message sent from $C_{1,m}$ to $V_m$ and the associated codeword bit, on each edge in the edge type $\mathcal {E}_{1m}$ connecting $C_{1,m}$ to $V_m$.

$I_{Ec}(2,m)$: The MI between the message sent from $C_{2}$ to $V_m$ and the associated codeword bit, on each edge in the edge type $\mathcal {E}_{2m}$ connecting $C_{2}$ to $V_m$.

Note that the extrinsic MI on an edge connecting $V_m$ to $C_{1,m}(\text{or}~C_{2})$, at the output of the variable node, is
the a-priori MI for $C_{1,m} (\text{or}~C_{2})$, i.e., $I_{Ev}(1,m)=I_{Ac}(1,m)~(\text{or}~I_{Ev}(2,m)=I_{Ac}(2,m))$. Similarly,
the extrinsic MI on an edge connecting $C_{1,m}~(\text{or}~C_{2})$ to $V_m$, at the output of the check node, is
the a-priori MI for $V_m$, i.e., $I_{Ec}(1,m)=I_{Av}(1,m)~(\text{or}~I_{Ec}(2,m)=I_{Av}(2,m))$. Then we have the iterative process as follows.

\textbf{$1.$ Variable nodes to check nodes update.}
The mean of the extrinsic MI on an edge type $\mathcal {E}_{1m}$ connecting $V_m$ to $C_{1,m}$, at the output of the variable node in the $l$-th iteration is
\begin{equation}\label{equ:19}
\begin{split}
&\phi_{V}^{(l)}(1,m)=\underset{d_{1,m=1}}{\overset{d_{v1,m}}{\boldsymbol\sum}}\underset{d_{2,m=0}}{\overset{d_{v2,m}}{\boldsymbol\sum}}
\left((d_{1,m}-1)\left[J^{-1}(I_{Av}^{(l)}(1,m))\right]^2+\right.\\&\left.d_{2,m}\left[J^{-1}(I_{Av}^{(l)}(2,m))\right]^2+\left[J^{-1}(I_{ch}(m))\right]^2\right)
\lambda_{[d_{1,m},d_{2,m}]}^{{1,m}}
\end{split}
\end{equation}
Also, the mean of the extrinsic MI on an edge of $\mathcal {E}_{2m}$ connecting $V_m$ to $C_{2}$, at the output of the variable node in the $l$-th iteration is
\begin{equation}\label{equ:20}
\begin{split}
&\phi_{V}^{(l)}(2,m)=\underset{d_{2,m=1}}{\overset{d_{v2,m}}{\boldsymbol\sum}}\underset{d_{1,m=1}}{\overset{d_{v1,m}}{\boldsymbol\sum}}
\left((d_{2,m}-1)\left[J^{-1}(I_{Av}^{(l)}(2,m))\right]^2+\right.\\&\left.d_{1,m}\left[J^{-1}(I_{Av}^{(l)}(1,m))\right]^2+\left[J^{-1}(I_{ch}(m))\right]^2\right)
\lambda_{[d_{1,m},d_{2,m}]}^{{2,m}}
\end{split}
\end{equation}
We can now get the MI in the $l$-th iteration as $I_{Ev}^l(1,m)=J\left(\sqrt{\phi_{V}^{(l)}(1,m)}\right)$ and $I_{Ev}^l(2,m)=J\left(\sqrt{\phi_{V}^{(l)}(2,m)}\right)$.

\textbf{$2.$ Check nodes to variable nodes update.} The update from check node to variable nodes is more complicated. We give the approximation according to~\cite{IEEEconf:13}.
The extrinsic MI on an edge type $\mathcal {E}_{1m}$ connecting $C_{1,m}$ to $V_m$, at the output of the check node in the $l$-th iteration is
\begin{equation}\label{equ:21}
\begin{split}
&I_{Ec}^{(l)}(1,m)=1-\\&J\left(\sqrt{\underset{d_{1,m}=1}{\overset{d_{c1,m}}{\boldsymbol\sum}}(d_{1,m}-1)\left[J^{-1}(1-I_{Ac}^{(l)}(1,m))\right]^2\rho_{[d_{1,m}]}^{1,m}}\right)
\end{split}
\end{equation}
The extrinsic MI on an edge type $\mathcal {E}_{2m}$ connecting $C_{2}$ to $V_m$ at the output of the check node in the $l$-th iteration is more complicated as
more than one source participates in the generation of $C_{2}$. We have
\begin{equation}\label{equ:22}
\begin{split}
&I_{Ec}^{(l)}(2,m)=1-\\&J\left(\sqrt{\underset{d_{2,m}=1}{\overset{d_{c2,m}}{\boldsymbol\sum}}\:\:\underset{\sim d_{2,m}}{\boldsymbol\sum}
\left((d_{2,m}-1)\left[J^{-1}(1-I_{Ac}^{(l)}(2,m))\right]^2+\right.}\right.
\\&\qquad\left.\overline{\underset{m'\neq m}{\underset{m'=1}{\overset{M}{\boldsymbol\sum}}}d_{2,l}\left[J^{-1}(1-I_{Ac}^{(l)}(2,m'))\right]^2\bigg)\rho_{[0,d_{2,1},\cdots,d_{2,M}]}^{2,m}}\right).
\end{split}
\end{equation}
In each iteration process, we make $I_{Av}(i,m)=I_{Ec}(i,m)$ and $I_{Ac}(i,m)=I_{Ev}(i,m)$ for $i=1,2$. $I_{ch}(m)$ is determined according to the SNR of
the $m$-th source-to-destination channel. At the end of the iteration, the MI between the $V_m$ and the associated codeword is
\begin{equation}\label{equ:23}
\begin{split}
&I(m)=J\left(\sqrt{\underset{d_{1,m=1}}{\overset{d_{v1,m}}{\boldsymbol\sum}}\underset{d_{2,m=0}}{\overset{d_{v2,m}}{\boldsymbol\sum}}
\left(d_{1,m}\left[J^{-1}(I_{Av}^{(l)}(1,m))\right]^2+\right.}\right.\\&\left.\overline{\left.d_{2,m}\left[J^{-1}(I_{Av}^{(l)}(2,m))\right]^2+\left[J^{-1}(I_{ch}(m))\right]^2\right)
\underset{i=1}{\overset{2}{\boldsymbol\sum}}\lambda_{[d_{1,m},d_{2,m}]}^{m}}\right)
\end{split}
\end{equation}
where
\begin{equation}\label{equ:24}
\lambda_{[d_{1,m},d_{2,m}]}^{{m}}=\frac{v_{[0,1],[d_{1,m},d_{2,m}]}(d_{1,m}+d_{2,m})}{\boldsymbol\sum_{d_{1,l=1}}^{d_{v1,m}}
\boldsymbol\sum_{d_{2,l=0}}^{d_{v2,m}}v_{[0,1],[d_{1,l},d_{2,l}]}(d_{1,l}+d_{2,l})}.
\end{equation}

According to the design of bilayer LDPC code, we first fix the lower graph codes of all sources, choosing the optimal point-to-point LDPC code to approach capacity $R_+^m$.
Then we optimize the overall graph to approach the capacity of the whole system.
The code optimization involves finding $M$ variable node degree distributions, i.e.,
$v_{[0,1],[d_{1,m},d_{2,m}]}$ for $s_m$ and a check node degree distribution $\mu_{[0,d_{2,1},\cdots,d_{2,M}]}$.
The optimization problem is concluded as minimizing the SNR of the whole system, which is a dual problem of
maximizing the system threshold. We can therefore write
\begin{equation}\label{equ:25}
\begin{split}
&\text{maximize} \quad \sigma_{sys}=\sqrt{\frac1{\frac1{(\sigma_-^1)^2}+\cdots+\frac1{(\sigma_-^M)^2}}}
\\&\text{subject to}~~ I(m)\rightarrow 1, ~~\text{for}~~ m=1,\cdots,M.
\end{split}
\end{equation}
\section{Numerical Results}\label{sec:5}
We choose the $2$-sources case to illustrate the design. The first source has the following capacities: $R_+^1=0.7,R_-^1=0.5$ and
$R_1^1=0.2$. The second source has the following capacities: $R_+^2=0.58,R_-^2=0.38$ and
$R_2^1=0.2$. Applying the proposed method, we get the code profile shown in Table~\ref{tab:1}.  
Note that $\sigma_-^1$ and $\sigma_-^2$ in Table~\ref{tab:1} are the thresholds for $R_+^1$ and $R_+^2$, respectively. 
Also note that the threshold deduced by EXIT is always larger than that deduced by DE, which has been proved by~\cite{IEEEconf:16}.
So $\sigma_-^1$ and $\sigma_-^2$ in the Table~\ref{tab:1} are larger than the exact values, which are searched by criterion~(\ref{equ:25}).
\begin{table}[tbh]
\centering
\begin{tabular}{|c|c|c||c|c|c|}
\hline
\multicolumn{3}{|c||}{\bfseries Source $1$} & \multicolumn{3}{|c|}{\bfseries Source $2$}\\
\hline
\hline
\multicolumn{6}{|c|}{\bfseries Variable Node Distribution}\\
\hline
$v_{[0,1][d_{1,1},d_{2,1}]}$ & $\mathcal {E}_{11}$ & $\mathcal {E}_{21}$ & $v_{[0,1][d_{1,2},d_{2,2}]}$ & $\mathcal {E}_{12}$ & $\mathcal {E}_{22}$\\
\hline
$0.244225$ & $2$ & $0$ & $0.3289203$ & $2$ & $0$\\
\hline
$0.1531552$ & $2$ & $1$ & $0.0772109$ & $2$ & $1$\\
\hline
$0.0209422$ & $2$ & $7$ & $0.0531292$ & $2$ & $2$\\
\hline
$0.1930021$ & $3$ & $0$ & $0.145309$ & $3$ & $0$\\
\hline
$0.138759$ & $3$ & $3$ & $0.0149215$ & $3$ & $1$\\
\hline
& & & $0.123802$ & $3$ & $2$\\
\hline
& & & $0.0286741$ & $3$ & $14$\\
\hline
$0.058304$ & $6$ & $0$ & $0.0346943$ & $6$ & $0$\\
\hline
$0.0109062$ & $6$ & $7$ & $0.0101216$ & $6$ & $2$\\
\hline
$0.000131148$ & $6$ & $21$ & & &\\
\hline
$0.05680712$ & $7$ & $0$ & $0.092595$ & $7$ & $0$\\
\hline
$0.047728$ & $7$ & $2$ & $0.0297043$ & $7$ & $7$\\
\hline
$0.0556525$ & $20$ & $3$ & $0.0165257$ & $20$ & $0$\\
\hline
$0.0203505$ & $20$ & $7$ & $0.00437642$ & $20$ & $1$\\
\hline
& & & $0.0400163$ & $20$ & $3$\\
\hline\hline
\multicolumn{6}{|c|}{\bfseries Check Node Distribution in Lower Graph}\\
\hline
\multicolumn{2}{|c|}{$\mu_{[d_{1,1}]}$} & $\mathcal {E}_{11}$ & \multicolumn{2}{|c|}{$\mu_{[d_{1,2}]}$} & $\mathcal {E}_{12}$\\
\hline
\multicolumn{2}{|c|}{$0.3$} & $15$ & \multicolumn{2}{|c|}{$0.42$} & $10$\\
\hline
\hline
\multicolumn{6}{|c|}{\bfseries Check Node Distribution in Upper Graph}\\
\hline
\multicolumn{2}{|c|}{$\mu_{[0,d_{2,1},d_{2,2}]}\times\frac{d_{2,1}}{d_{2,1}+d_{2,2}}$} & $\mathcal {E}_{21}$ & \multicolumn{2}{|c|}{$\mu_{[0,d_{2,1},d_{2,2}]}\times\frac{d_{2,2}}{d_{2,1}+d_{2,2}}$} & $\mathcal {E}_{22}$\\
\hline
\multicolumn{2}{|c|}{$0.4\times 0.5$} & $3$ & \multicolumn{2}{|c|}{$0.4\times 0.5$} & $3$\\
\hline
\hline
\multicolumn{6}{|c|}{\bfseries Code Rate and Threshold}\\
\hline
\multicolumn{3}{|c||}{$R_+^1=0.7,~R_-^1=0.5$} & \multicolumn{3}{|c|}{$R_+^2=0.58,~R_-^2=0.38$}\\
\hline
\multicolumn{3}{|c||}{$\sigma_+^1=0.722955,~\sigma_-^1=0.970555$} & \multicolumn{3}{|c|}{$\sigma_+^2=0.859273,~\sigma_-^2=1.189900$}\\
\hline
\end{tabular}
\caption{The node distribution of the NCMET-LDPC in two-source case.}\label{tab:1}
\end{table}

First of all, we determine the distributions of $\mathcal {E}_{11}$ and $\mathcal {E}_{12}$ in the lower graphs for $s_1$ and $s_2$
to approach the rate $R_+^1$ and $R_+^2$, respectively. They are designed as the single link LDPC codes and can be directly obtained from~\cite{IEEEconf:15}.
Our main task is to find out the optimal distribution of $\mathcal {E}_{21}$ and $\mathcal {E}_{22}$, and the corresponding $v_{[0,1][d_{1,1},d_{2,1}]}$,
$v_{[0,1][d_{1,2},d_{2,2}]}$ and $\mu_{[0,d_{2,1},d_{2,2}]}$. By adopting the searching criterion of (\ref{equ:25}) and EXIT curves fitting,
we get the three elements with the thresholds $\sigma_-^1=0.970555$ and $\sigma_-^2=1.1899$ for $s_1$ and $s_2$, respectively.
The variable node distributions of the two sources, $v_{[0,1][d_{1,1},d_{2,1}]}$, $v_{[0,1][d_{1,2},d_{2,2}]}$, and corresponding degrees are shown in Table~\ref{tab:1}.
The check node distribution in the upper graph at the relay is $\mu_{[0,d_{1,m},d_{2,m}]}$, which has only one distribution as $\mu_{[0,d_{1,m},d_{2,m}]}=R_-^1+R_-^2=0.4$.
Each check node of the relay has the degree $6$, half of which are allocated to $s_1$ and the other half are allocated to $s_2$.
This is reasonable since $R_-^1=R_-^2=0.2$.

\begin{figure}[!t]
\centering
\includegraphics[width=3.5in,angle=0]{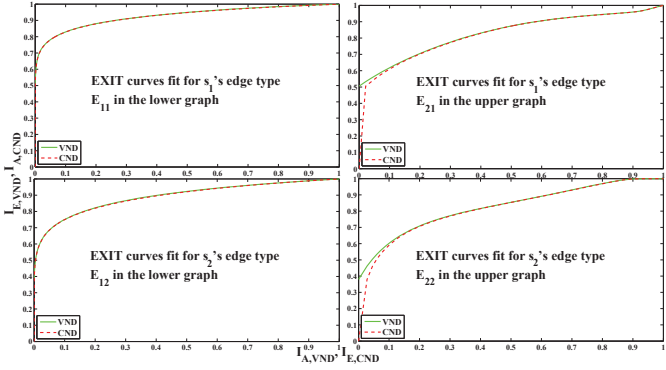}
\caption{EXIT curves for two sources at the system threshold $\sigma_-^1=0.970555$ with rate $R_-^1=0.5$
and $\sigma_-^2=1.1899$ with rate $R_-^2=0.38$.}\label{fig6}
\end{figure}

\begin{figure}[!t]
\centering
\includegraphics[width=3.5in,angle=0]{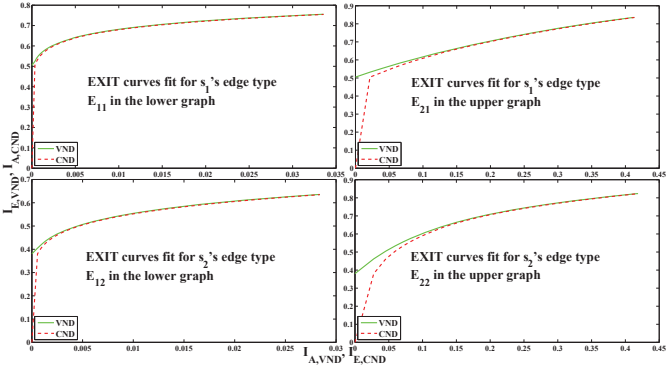}
\caption{EXIT curves with $\sigma_-^1=0.969555$ \text{and} $\sigma_-^2=1.1909$.}\label{fig7}
\end{figure}

\begin{figure}[!t]
\centering
\includegraphics[width=3.5in,angle=0]{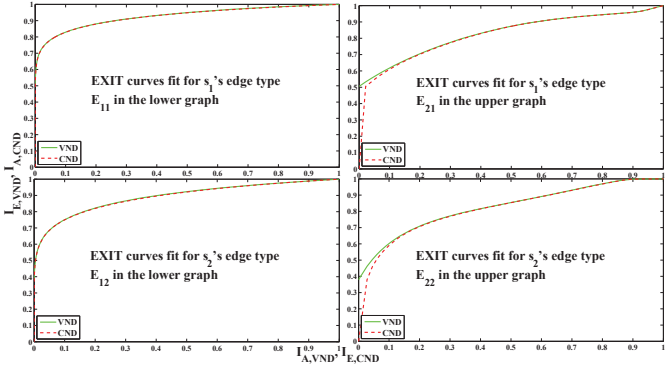}
\caption{EXIT curves with $\sigma_-^1=0.971555$ \text{and} $\sigma_-^2=1.1889$.}\label{fig8}
\end{figure}

\begin{figure}[!t]
\centering
\includegraphics[width=3.8in,angle=0]{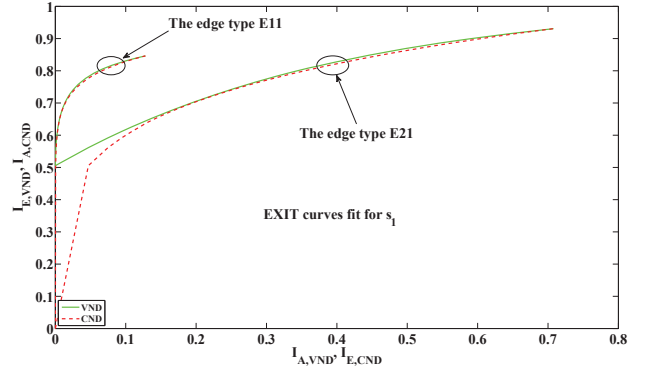}
\caption{EXIT curves fit for two sources at the system threshold $\sigma_-^1=0.970555$ with rate $R_-^1=0.5$
and $\sigma_-^2=1.1899$ with rate $R_-^2=0.38$.}\label{fig9}
\end{figure}

Fig.~\ref{fig6} shows the EXIT charts for the $4$ edge types at the system threshold.
Fig.~\ref{fig7} and Fig.~\ref{fig8} show the EXIT chart at $\sigma_-^1=0.969555\text{,}~\sigma_-^2=1.1909$ and
$\sigma_-^1=0.971555\text{,}~\sigma_-^2=1.1889$, respectively. Note that in the three figures, we adopt the same value of the summation of
$\sigma_-^1$ and $\sigma_-^2$. However, the system threshold in Fig.~\ref{fig7} is larger than that of Fig.~\ref{fig6},
and the system threshold in Fig.~\ref{fig8} is smaller than that of Fig.~\ref{fig6}. So we can see that the EXIT curves
in Fig.~\ref{fig7} intersect at a value smaller than $1$, which means that the iterative decoding at destination will not converge eventually.
The EXIT curves in Fig.~\ref{fig8} as well as Fig.~\ref{fig6} intersect at $1$, which means the decoding at destination will succeed. Finally, we investigate the separate LDPC code without network coding. In this case, each source is only connected to half of the check nodes at the relay, and each check node at relay only contains one edge type, either $\mathcal E_{2,1}$ or $\mathcal E_{2,2}$. On the other hand, we keep the variable node distributions of each source unchanged. Fig.~\ref{fig9} shows the EXIT curves of the two edge types connected to $s_1$. Obviously, both edge types cannot converge at $1$. So we conclude from this case that the separate LDPC code obtains less extrinsic mutual information from the check nodes at relay than the proposed coding scheme.

\section{Conclusion}
In this paper, we investigate a network coded LDPC design in the multi-source scenario.
We apply the multi-edge LDPC to the system and execute the EXIT analysis. We conclude that
each source achieves more extrinsic mutual information due to joint processing at the relay.
Therefore, our scheme delivers better performance  compared to  traditional schemes that do not utilize joint processing at the relay.


%


\end{document}